# The Nash Equilibrium Revisited: Chaos and Complexity Hidden in Simplicity[1]

**Philip Vos Fellman**
Southern New Hampshire University
Manchester, NH
Shirogitsune99@yahoo.com

[1] The author gratefully acknowledges the assistance of the Santa Fe Institute in providing the materials from which the majority of this paper has been drawn. In particular, thanks are due to Thomas Kepler, Vice President for Academic Affairs and J. Doyne Farmer, whose theoretical breakthroughs and continuing efforts to bridge the disciplinary gap between physics and finance have helped to make this paper possible. Dr. Massood Samii, Chairman of the International Business Department and Dr. Paul Schneiderman, Dean of he School of Business, Southern New Hampshire University generously supported this research through release time and travel expenses as well.

## Introduction

The Nash Equilibrium is a much discussed, deceptively complex, method for the analysis of non-cooperative games.[2] If one reads many of the commonly available definitions the description of the Nash Equilibrium is deceptively simple in appearance. Modern research has discovered a number of new and important complex properties of the Nash Equilibrium, some of which remain as contemporary conundrums of extraordinary difficulty and complexity. Among the recently discovered features which the Nash Equilibrium exhibits under various conditions are heteroclinic Hamiltonian dynamics,[3] a very complex asymptotic structure in the context of two-player bi-matrix games[4] and a number of computationally complex or computationally intractable features in other settings.[5] This paper reviews those findings and then suggests how they may inform various market prediction strategies.

One of the particularly difficult aspects of the Nash Equilibrium is that the average economic practitioner, game theorist or other type of modeler is most likely to be familiar with a very truncated simplified version of John Nash's actual discovery. In business schools, for example, phrases like "a Nash equilibrium occurs whenever you have a strategy which cannot be dominated by any other strategy but which itself cannot dominate any other strategy". Another, somewhat more elaborate explanation[6] is given by Brandeis University's computer science department:[7]

> A Nash equilibrium (formulated by John Nash ) is best described in words as a set of best strategies chosen by one player in response to the strategies chosen by the other players. Axelrod gives a mathematical definition of a Nash equilibrium of a game in terms of the normal form of the game. Gintis talks about a specific kind of Nash equilibria called a pure strategy Nash equilibrium where every player uses a "pure" strategy. Gintis doesn't say what he means by a "pure strategy," stating only that a pure strategy is not necessarily a dominant strategy, and then presents many examples of games where one can find a pure Nash equilibrium (but he leaves it up to the reader to figure out what that Nash equilibrium is), such as chess. (It is my guess that a pure strategy is one where the payoffs are the highest, but the dominant strategy may not lead to the best payoffs). In fact, any game with perfect information (such as chess, because two players move, but there are no outside factors such as Nature), has a pure strategy Nash equilibrium (proved by Harold Kuhn in 1953).

In a scientific sense, all of these simplified conceptions, many of which are very useful heuristics or generalizations, really dance around the nature, structure and meaning of the Nash Equilibrium. While a

---

[2] (A) McLennan, Andrew and Berg, Johannes, "The Asymptotic Expected Number of Nash Equilibria of Two Player Normal Form Games argue that "Few game theorists would disagree with the assertion that, at least subjectively, Nash Equilibrium is a complex concept. For certain particular classes of games it is possible to analyze the set of Nash Equilibria without undue difficulty, but even seemingly quite minor modifications of a game from such a class typically render equilibrium analysis intractable. For general extensive or normal form games, only small examples can be solved by hand. New software has significantly extended the range of games for which equilibrium analysis is possible, but it is still easy to bump up against the limitations stemming from computational complexity."

John Milnor recounts "John Forbes Nash published his first paper with his father at age seventeen. His thesis, at age twenty-one, presented clear and elementary mathematical ideas that inaugurated a slow revolution in fields as diverse as economics, political science, and evolutionary biology. During the following nine years, in an amazing surge of mathematical activity, he sought out and often solved the toughest and most important problems he could find in geometry and analysis. Then a mental breakdown led to thirty lost and painful years, punctuated by intermittent hospitalization, as well as occasional remission. However, in the past ten years a pronounced reawakening and return to mathematics has taken place. Meanwhile, the importance of Nash's work has been recognized by many honors: the von Neumann Prize, fellowship in the Econometric Society and the American Academy of Arts and Sciences, membership in the U.S. National Academy of Sciences, culminating in a Nobel Prize." Notices of the American Mathematical Society 1329, Volume 45, No. 10, November, 1998.

[3] Yuzuru Sato, Eizo Akiyama and J. Doyne Farmer, (2001) "Chaos in Learning a Simple Two Person Game", Santa Fe Institute Working Papers, 01-09-049. Subsequently published in Proceedings of the National Academy of. Sciences,. USA, 99, pp. 4748-4751, (2002).

[4] See, for example Quint, Thomas and Shubik, Martin, (1997) "A Bound on the Number of Nash Equilibria in a coordination game", Cowles Foundation Discussion Paper 1095, Yale University, 1997 at
http://cowles.econ.yale.edu/P/cd/d10b/d1095.pdf

[5] See for example, Pang, Jong-Shi and Fukushima, Masao "Quasi-Variational Inequalities, Generalized Nash Equilibria and Multi-Leader-Follower Games," at www.mts.jhu.edu/~pang/fupang_rev.pdf where the authors note in describing the use of the Nash Equilibrium for utility pricing that "the bidding strategy of the dominant firm (the leader) in a competitive electricity market, whose goal is to maximize profit subject to a set of price equilibrium constraints… is a difficult, non-convex optimization problem ; efficient methods for computing globally optimal solutions are to date not available. In practice the multi-dominant firm problem is of greater importance; and yet, as a Nash game, the latter can have no equilibrium solution in the standard sense." Technical Report 2002-009, Department of Applied Mathematics and Physics, Kyoto University (September 2002).

[6] http://www.cs.brandeis.edu/~cs113/classprojects/~earls/cs113/WP1.html

[7] A more thorough treatment of co-ordination games is given by Roger McCain, of Drexel University at http://william-king.www.drexel.edu/top/eco/game/nash.html

detailed exploration of Nash's discovery is beyond the scope of the current paper, it is recommended that the interested reader download Nash's doctoral dissertation, "Non-Cooperative Games", which is the 28 page mathematical proof that constitutes the actual "Nash Equilibrium".[8]

Research on the Nash Equilibrium in its complete form is generally highly abstract and often involves a degree of complexity which makes that research rather daunting for anyone but the professional mathematician or mathematical economist. A good example of the complexity of research which has added valuable extensions to our knowledge of the Nash Equilibrium in a technical sense is the series of papers (in some sense competitive and in another sense incrementally building on the work of its predecessors) concerning asymptotic boundaries for Nash Equilibria in particular types of game-theoretic situations.

## Asymptotic Boundaries for Nash Equilibria in Non-Cooperative Games

Thomas Quint and Martin Shubik began their exploration of this element of the Nash Equilibrium in their 1997 paper, "A Bound on the Number of Nash Equilibria in a Coordination Game".[9] Quint and Shubik's abstract states:[10]

> We show that if y is an odd integer between 1 and $2^n - 1$, there is an $n \times n$ bimatrix game with exactly $y$ Nash equilibria (NE). We conjecture that this 2n - 1 is a tight upper bound on the number of NEs in a "nondegenerate" n x n game. We prove the conjecture for $N \leq 3$, and provide bounds on the number of NE's in $m \times n$ nondegenerate games when min $(m,n) \leq 4$.

For those not professionally familiar with this type of proof, the subject is made more accessible in the main body of the text, where Quint and Shubik argue that:[11]

> In two-person game theory, perhaps the most important model is the so-called "strategic form" or "bi-matrix" game form of a game. The game is represented as a two-dimensional matrix, in which rows represent the (pure) strategies for one player and the columns those for the other. In each cell is placed a pair of numbers representing the payoffs to the two players if the corresponding pair of strategies is chosen.
> 
> In the analysis of bi-matrix games, perhaps the most basic solution concept is that of Nash Equilibrium (NE). A pair of (mixed) strategies (p*, q*) is an NE provided the first player cannot do any better than to play p* against the second player's q*, while likewise, the second player's "best response" against p* is q*. In Nash's seminal papers (1950, 1953), he proved that every bimatrix game has a NE mixed strategy.

Notice that this is a rather more precise definition of the concept of a strategy which cannot be dominated but which also cannot dominate other strategies in the game.

Quint and Shubik's paper is important for a number of a reasons. First, they provide a succinct explanation of the work done over the past fifty years on studying the Nash Equilibrium. In particular, they capture the essence of the problems inherent in defining the number of NE's in a bi-matrix game. For those who have not followed this research, John Nash's original work (1950)[12] says that there has to be at least one "Nash Equilibrium" in such a game (Theorem 1, p. 5). In 1964, Lemke and Howson[13] demonstrated that under a certain "nondegeneracy" assumption, the number of NE's must be finite and odd. Quint and Shubik note that this "oddness" property appears in various forms over the years, and then in 1993, Gul, Pearce and Staccheti demonstrated that if a nondegenerate game has 2y-1 NE's, at most, y of them are pure strategy NE's. Shapley argued as early as 1974 that in a 3 x 3, non-degenerate game the maximum number of NE's was seven, but provided no proof.[14] Quint and Shubik also note that the situation is less clear for cases of N > 3. They argue that it is an open problem to fully characterize the numbers of NE that can occur and offer a partial solution in their 1997 paper.[15]

---

[8] http://www.princeton.edu/mudd/news/faq/topics/Non-Cooperative_Games_Nash.pdf
[9] http://cowles.econ.yale.edu/P/cd/d10b/d1095.pdf
[10] Ibid.
[11] Ibid.
[12] Ibid. No. 7
[13] Lemke. C.E. and Howson, J.T. (1964) "Equilibrium points of bimatrix games", Journal of the Society for Industrial and Applied Mathematics, 12:413--423, 1964.
[14] Shapley, L.S. (1974) "A note on the Lemke-Howson Algorithm", Mathematical Programming Study 1: Pivoting and Extensions, 175-189 (1974).
[15] Ibid. No. 10

Perhaps the most interesting aspect of Quint and Shubik's paper was that even while they were significantly extending knowledge about the Nash Equilibrium, they were too modest in their conjecture. Almost immediately following the Quint and Shubik paper, Bernhard von Stengel of the London School of Economics in a paper entitled "New lower bounds for the number of equilibria in bimatrix games" argued that:[16]

> A class of nondegenerate n \Theta n bimatrix games is presented that have asymptotically more than 2:414 n = p n Nash equilibria. These are more equilibria than the 2 n \Gamma 1 equilibria of the game where both players have the identity matrix as payoff matrix. This refutes the Quint--Shubik conjecture that the latter number is an upper bound on the number of equilibria of nondegenerate n \Theta n games. The first counterexample is a 6 \Theta 6 game with 75 equilibria…

Subsequently, At the First World Congress of the Game Theory Society[17] von Stengel presented an improved computational method for finding the equilibria of two-player. His method was based on what he describes as the "*sequence form* of an extensive form game with perfect recall". Von Stengel described this method as "improvements of computational methods for finding equilibria of two-player games in extensive form", which was "based on the 'sequence form' of an extensive form game with perfect recall, a concept published by the author [*von Stengel*] in GEB in 1996, and in parts independently described by other authors (Romanovskii 1961, Selten 1988, Koller and Megiddo 1992)." The brunt of von Stengel's presentation was to "show a number of results that demonstrate how to put this result into practice, in particular to achieve numerical stability." This method, developed around algorithms for polytope computing, created algorithms which were capable of computing what von Stengel subsequently describes as "single equilibria that are (strategic-form) trembling-hand perfect and, for smaller games, allow for the complete enumeration of all equilibrium components."[18] From a purely computational point of view, this result is a very significant theoretical advance in the understanding and ability to calculate the complex structure of the Nash Equilibrium.

Credit for the final "blockbuster" on this subject shout go to Andrew McLennan and In-Uck Park, for their 1999 paper, ""Generic 4×4 Two Person Games Have at Most 15 Nash Equilibria".[19] Here, the authors used an articulated cyclic polytope model to construct a series of lemmas that constrain the set of equilibria. Polytope computing, like quantum computing is one of the more powerful non-linear solutions which has been suggested for dealing with both this and other, recently discovered structural properties of the Nash Equilibrium which are, for contemporary, digital computing architectures, computationally intractable.

---

[16] von Stengel, Bernhard (1997) "New lower bounds for the number of equilibria in bimatrix games" Technical Report 264, Dept. of Computer Science, ETH Zurich, 1997. http://citeseer.nj.nec.com/vonstengel97new.html
[17] von Stengel, Bernhard (2000) "Improved equilibrium computation for extensive two-person games", First World Congress of the Game Theory Society (Games 2000), July 24-28, 2000 Basque Country University and Fundacion B.B.V., Bilbao, Spain.
[18] Ibid.
[19] McLennan, A and Park, I (1999) "Generic 4×4 Two Person Games Have at Most 15 Nash Equilibria", Games and Economic Behavior, 26-1, (January, 1999), 111-130.

# Non-Cooperative Meta-Game Structures: The Rationality Trap

In a 2002 paper, Sato Yuzuru, Akiyama Eizo and J. Doyne Farmer present a novel view of the Nash equilibrium in the case of learning to play a simple, two player, rock-paper-scissors game. Their abstract argues rather modestly:[20]

> We investigate the problem of learning to play a generalized rock-paper-scissors game. Each player attempts to improve her average score by adjusting the frequency of the three possible responses. For the zero-sum case the learning process displays Hamiltonian chaos. The learning trajectory can be simple or complex, depending on initial conditions. For the non-zero-sum case it shows chaotic transients. This is the first demonstration of chaotic behavior for learning in a basic two person game. As we argue here, chaos provides an important self-consistency condition for determining when adaptive players will learn to behave as though they were fully rational.

While this sounds relatively straightforward, the actual constraints of rationality, whether the strong "instrumentally rational utility maximization" of neoclassical economics"[21] or the more limited "bounded rationality" of neo-institutionalism[22] and rational choice theory,[23] have strong effects on the decision process and its end points.

The pioneer in this field is W. Brian Arthur, developer of modern non-equilibrium economics. Without attempting to recapitulate this very large body of work,[24] it is useful to look at a few key ways in which equilibrium assumptions about transactions, and strategies, when based on assumptions about the rationality of play on the part of the opposing players, encounter a great deal of difficulty. To the extent that these strategies (including strategies which attempt to find a Nash equilibrium) dictate one player's choices by attempting to anticipate the choices or strategies of other players, they are not robust and their fragility can easily cause the entire game to become ill-defined (in economic terms this would be defined as "market failure"). As Arthur himself argues:[25]

> The type of rationality we assume in economics--perfect, logical, deductive rationality--is extremely useful in generating solutions to theoretical problems. But it demands much of human behavior--much more in fact than it can usually deliver. If we were to imagine the vast collection of decision problems economic agents might conceivably deal with as a sea or an ocean, with the easier problems on top and more complicated ones at increasing depth, then deductive rationality would describe human behavior accurately only within a few feet of the surface. For example, the game Tic-Tac-Toe is simple, and we can readily find a perfectly rational, Minimax solution to it. But we do not find rational "solutions" at the depth of Checkers; and certainly not at the still modest depths of Chess and Go.

As a concrete example of how anticipatory strategies can rapidly degenerate, Arthur offers the following example:

> There are two reasons for perfect or deductive rationality to break down under complication. The obvious one is that beyond a certain complicatedness, our logical apparatus ceases to cope--our rationality is bounded. The other is that in interactive situations of complication, agents can not rely upon the other agents they are dealing with to behave under perfect rationality, and so they are forced to guess their behavior. This lands them in a world of subjective beliefs, and subjective beliefs about subjective beliefs. Objective, well-defined, shared assumptions then cease to apply. In turn, rational, deductive reasoning--deriving a conclusion by perfect logical processes from well-defined premises--itself cannot apply. The problem becomes ill-defined.

Two classic examples of his kind of strategic mis-fire which are particularly interesting are Arthur's "El Farol Bar Game", and Epstein and Hammond's "line-up game". The El Farol Bar problem originally presented by Arthur was been formally elaborated as "The Minority Game" by Damien Challet and Yi-Cheng Zhang (See appendix I). In this game the structure is such that there there is no equilibrium point which can mutually satisfy all players. This is not because the equilibrium is unattainable due to practical constraints but rather because in the manner described above, each player's move changes the expectations and payoffs of the other players (a feedback loop not terribly dissimilar to some of Arthur's path-dependent stochastic processes).[26] In Epstein and Hammond's "Non-explanatory Equilibria: An Extremely Simple Game With (Mostly) Unattainable Fixed Points" the equilibrium points of the game are so sparsely distributed (i.e., the phase space is so large) that even small increases in the number of players creates a solution basin of such high dimensionality that the problem becomes not just ill-defined but insoluble.[27] Both of these cases involve complex structures where small, linear growth in one parameter drives exponential growth in another parameter, with that exponential growth often feeding back into the original iterative level of the process.[28]

---

[20] Ibid., No. 3
[21] See, for example, (a) Cox, Gary W. (2001) "Lies, damned lies, and rational choice analyses" in <u>Problems and Methods in the Study of Politics</u>, Cambridge University Press, (forthcoming, 2004).
[22] See, for example, Knott, Jack and Miller, Gary (1988) <u>Reforming Bureaucracy</u>, Prentice-Hall, 1988.
[23] See Shepsle, Kenneth A., and Bonchek, Mark S., <u>Analyzing Politics: Rationality, Behavior, and Institutions</u>, W.W. Norton, 1997.
[24] See (a) Arthur, W. Brian, <u>Increasing Returns and Path Dependence in the Economy</u>, University of Michigan Press, 2nd Edition, 1994; (b) and his multiple book series The Economy As an Evolving Complex System, 1988-1997 (Santa Fe Institute Studies in the Science of Complexity).
[25] Arthur, W. Brian "Inductive Reasoning and Bounded Rationality", American Economic Review, (Papers and Proceedings), 84,406-411, 1994.

## Chaos in Learning a Simple Two Person Game:

The Sato, Akiyama and Farmer paper (referred to hereafter as the SAF paper or the SAF team), starts off with the argument that because of bounded rationality, players in even a relatively simple game like rock-paper-scissors, may not, in fact, converge to the Nash equilibrium (which is the expectation for rational or instrumentally rational players).[29] Given the problem of non-convergence, they argue that when players fail to learn the Nash equilibrium strategy in a multiple iteration or multiple generation game, it then becomes important to understand what the actual dynamics of the learning process in that situation are. For analytical purposes they treat the learning "trajectory" as the phase space representation of the degree to which player's game choices (strategies and moves) either become asymptotic to the Nash equilibrium or fail to converge upon the Nash Equilibrium at all.

What makes the SAF paper so interesting is the authors' novel insight into the mechanics of the zero-sum rock-paper-scissors game. Having recognized that one player's loss is another player's gain, they proceed to demonstrate how the conserved quantity of the payoff endows the learning dynamics of the game with a Hamiltonian structure resembling that of celestial mechanics.[30] While celestial mechanics comprises one of the most deterministic systems in physics, it is also extraordinarily complex. The SAF team notes that in the context of studying the learning trajectories of players ostensibly moving towards the Nash equilibrium:[31]

> Because of the zero sum condition the learning dynamics have a conserved quantity, with a Hamiltonian structure similar to that of physical problems such as celestial mechanics. There are no attractors and trajectories do not approach the Nash Equilibrium. Because of the Hamiltonian structure the chaos is particularly complex, with chaotic orbits finely interwoven between regular orbits; for an arbitrary initial condition it is impossible to say a priori which type of behavior will result. When the zero sum condition is violated we observe other complicated dynamical behaviors such as heteroclinic orbits with chaotic transients. As discussed in the conclusions, the presence of chaos is important because it implies that it is not trivial to anticipate the behavior of the other player. Thus under chaotic learning dynamics even intelligent, adapting agents may fail to converge to a Nash Equilibrium.

---

[26] Explained in Farmer, J. Doyne "Physicists Attempt to Scale the Ivory Towers of Finance", Computing in Science and Engineering, November-December, 1999.
[27] Epstein, Joshua M, and Hammond, Ross A "Non-Explanatory Equilibria: An Extremely Simple Game with (Mostly) Unattainable Fixed Points." in Complexity, Vol.7, No. 4: 18-22, 2002.
[28] Martin Shubik and Nicholas J. Vriend, for example, argue in "A Behavioral Approach to a Strategic Market Game" (Santa Fe Institute, 1998) that "The size and complexity of the strategy sets for even a simple infinite horizon exchange economy are so overwhelmingly large that it is reasonably clear that individuals do not indulge in exhaustive search over even a large subset of the potential strategies. Furthermore, unless one restricts the unadorned definition of a noncooperative equilibrium to a special form such as a perfect noncooperative equilibrium, almost any outcome can be enforced as an equilibrium by a sufficiently ingenious selection of strategies. In essence, almost anything goes, unless the concept of what constitutes a satisfactory solution to game places limits on expected behavior."
[29] Ibid., No. 3, "A failure to converge to a Nash equilibrium under learning can happen, for example, because the dynamics of the trajectories of the evolving strategies in the space of possibilities are chaotic. This has been observed in games with spatial interactions, or in games based on the single population replicator equation . In the latter examples players are drawn from a single population and the game is repeated only in a statistical sense (i.e. the identity of the players changes in repeated trials of the game)." (p. 1)
[30] Ibid., No. 3
[31] Ibid., No. 3

# Modeling Learning Behavior for Nash Equilibrium Problems with Conserved Payoff Quantities – Regularity, Chaos and Chaotic Transients

Not surprisingly, replicator equations have played a significant role in the SAF team's research. They approach the subject by first noting that for some time it has been well known that chaos occurs in single population replicator equations. They then apply this to game theory in the specialized context where both players are forced to use the same strategy, giving the example of two statistically identical players being repeatedly drawn from the same population. This process is then applied to the playing of a game with two fixed players who evolve their strategies independently. The generalized dynamics of this system are shown in Appendix II. In an exemplification of their findings, the SAF team then models a real-life competitive scenario where no strategy is dominant and there is no pure strategy Nash Equilibrium.

For their example they use two broadcasting companies competing for the same time slot when the preferences of the audience are context dependent. They then ask the following question: "Suppose, for example, that the audience prefers sports to news, news to drama and drama to sports."[32] If each broadcasting company must commit to their schedule without knowing that of their competitor, then the resulting game is of the type shown in Appendix II.[33]

The next step in the analysis is to show that the example given above fits the general case, where a tie is not equivalent for both players.[34] The argument here is that no strategy dominates as long as the audience believes that within any given category one company's programming is superior to the other. Assuming that we are still looking at a zero sum game, then the size of the audience must be fixed and one company's gain will be the other company's loss.[35] In this case the coupled replicator equations form a conservative system *which cannot have an attractor*. The implications of this relationship are profound.[36] In the most critical case, the system becomes non-integrable and as chaotic trajectories extend deeper into the strategy space, regular and chaotic trajectories become so finely interwoven that *there is a regular orbit arbitrarily close to any chaotic orbit*. To exchange sides of the equation, calculationally, as values of ϵ approach 0.5, any regular orbit is arbitrarily close to a chaotic orbit. Not only is this in certain very

---

[32] Economically sophisticated readers will immediately recognize this as the Arrow Paradox, named after Nobel Laureate, Kenneth Arrow. Those wishing to see a complete treatment of the Arrow Paradox are recommended to John Geanakoplos, "Three Brief Proofs of Arrow's Impossibility Theorem", Cowles Foundation Discussion Paper No. 1123RRR, The Cowles Foundation for Research in Economics, Yale University, June 2001, available at http://cowles.econ.yale.edu/P/au/d_gean.htm

[33] Ibid. No. 3.

[34] i.e., ϵx ≠ ϵy

[35] Ibid. No. 3: Mathematically this corresponds to the condition ϵx = – ϵy. The structure of the game is such that at each move the first player chooses from one of m possible pure strategies (moves) with a frequency of x = ($x_1, x_2, \ldots x_m$) and similarly the second player chooses from one of n possible pure strategies with a frequency of y = ($y_1, y_2, \ldots y_n$). The players update x and y based on past experience using reinforcement learning. Behaviors that have been successful are reinforced and those that have been unsuccessful are repressed. In the continuous time limit where the change in x and y on any given time step goes to zero, under some plausible assumptions it is possible to show that reinforcement dynamics are described by coupled replicator equations of the form (p. 2):

$\dot{x}_i = x_i [(Ay)i – \mathbf{x}AY]$, (Equation 1)

$\dot{y}_i = y_i [(Bx)I – \mathbf{y}Bx]$ (Equation 2)

[36] Ibid. (p. 2):

"Since in this case Aij = -Bji, it is known that the dynamics are Hamiltonian. This is a stronger condition, as it implies the full dynamical structure of classical mechanics, with pairwise conjugate coordinates obeying Liouville's theorem (which states that the area in phase space enclosed by each pair of conjugate coordinates is preserved)."

There follows a complex transformation of coordinates, giving the Hamiltonian of:

$H = -1/3 (u_1 + u_2 + v_1 = v_2) + \log (1 + e^{u1} + e^{u2})(1 + e^{v1} + e^{v2})$

$$J = \begin{bmatrix} 0 & 0 & 2\epsilon & 3+\epsilon \\ 0 & 0 & -3+\epsilon & 2\epsilon \\ -2\epsilon & 3-\epsilon & 0 & 0 \\ -3-\epsilon & -2\epsilon & 0 & 0 \end{bmatrix}$$

$\dot{U} = J\nabla_U H$ where the Poisson structure **J** is given as:

The original working paper posted on the worldwide web gives three dimensional dynamic graphics of Poincare sections of equations 1 and 2. It can be downloaded from http://www.santafe.edu/sfi/publications/Abstracts/01-09-049abs.html The diagrams show a sample of points where the trajectories intersect the hyperplane $x_2 – x_1 + y_2 – y_1 = 0$. "When ϵ = 0, the system is integrable and trajectories are confined to quasi-periodic torii, but when ϵ > 0 this is no longer guaranteed. As we vary ϵ from 0 to 0.5, without changing initial conditions, some torii collapse and become chaotic, and the trajectories cover a larger region of the strategy space. **Regular and chaotic trajectories are finely interwoven; for typical behavior of this type, there is a regular orbit arbitrarily close to any chaotic orbit**." (p. 2) fundamental ways computationally intractable, but it makes a shambles of any predictive methodology for system trajectories. The SAF group then calculated Lyapunov exponents, which they treat as "the generalizations of eigenvalues that remain well defined for chaotic dynamics."[37] The graphic display of λ values which fluctuate narrowly when the tie breaking parameter $\epsilon = 0$, show dramatically increasing elements of chaotic fluctuation at $\epsilon = .25$ and even more so at the maximum value of $\epsilon = 0.5$[38]

## Surprising Results

An interesting conclusion of the SAF group is that this game has the possibility for an interesting mixed strategy Nash equilibrium when all responses are equally likely:
$x^*_1 = x^*_2 = x^*_3 = y^*_1 = y^*_2 = y^*_3 = 1/3$. While the authors argue that it is possible to show that the average of all trajectories has the same payoff as the Nash Equilibrium, the deviations from this payoff are so significant on any particular given orbit (and remember that as $\epsilon > 0.5$ any regular orbit is arbitrarily close to a chaotic orbit, rational, risk-averse
 agents, were they able to express the preference, would prefer the Nash Equilibrium.

The SAF team found another interesting and unexpected result in the learning patterns when their trajectories approached the heteroclinic pattern in the zero sum game.[39] The first interesting result is that players switched in a well defined order among their strategies varying play from rock to paper to scissors. Players were found under these conditions to spend time near each pure strategy with frequency increasing linearly time.[40]

An interesting contrast was that with a moderate change in variables ($\epsilon_x + \epsilon_y > 0$) the behavior is similar from a psychological point of view (the players gravitate increasingly towards pure strategies over time) however the mathematical behavior of the players is profoundly different. In the first case, the trajectory is attracted to the heteroclinic cycle and while there are sharp peaks, they increase in a regular, linear fashion.[41] In the second case the trajectory is a chaotic transient attracting to a heteroclinic orbit at the boundary of the simplex.[42]

This is chaotic behavior in its strong form even in the simplest of games. As the SAF team remarks, *"the emergence of chaos in such a simple game demonstrates that rationality may be an unrealistic approximation even in elementary settings."*[43] The authors illustrate how this kind of problem (chaos) in even simple game theoretic behavior may influence the behavior of agents, shedding some new light on the principal-agent problem. They contrast the situation of regular learning with that characterized by chaotic transients by examining agent behavior under both conditions. In the first case, they argue that even the simplest or least competent of agents can exploit her adversary's learning by even the crudest methods of approximation to improve her own performance. This is more or less the classical assumption underlying evolutionary strategies in multiple iteration games or agent exchanges.[44] Interestingly, the authors argue that chaos is actually an important *self-consistency* condition (although this might be an intuitive proposition for the experienced gamer, bargainer or negotiator, it certainly is not what drives the exceedingly complex mathematics of the learning dynamics in these simple games.)[45] Following this line of argument, they contend that when the learning of a competing agent is too regular, it is too easy for an opponent to exploit this behavior by simple extrapolation. On the other hand, having seen how far the mathematical complexity of chaos extends in even simple games, we can let the authors go ahead and guess the outcome which their paper proves, which is simply that "when behavior is chaotic…extrapolation is difficult, even for intelligent humans."[46] As the authors have already noted in detail, the Hamiltonian dynamics characteristic of this simple game situation are "particularly complex due to the lack of attractors and the fine interweaving of regular and irregular motion." They further note that "this situation is compounded for high dimensional chaotic behavior, due to the *curse of dimensionality*. In dimensions greater than about five, the amounts of data an *economic* agent would need to collect to build a reasonable model to extrapolate the learning behavior of the opponent becomes enormous. For games with more players it is possible to extend the replicator framework to systems of arbitrary dimension."[47] The SAF group concludes with the observation that "it is striking that low dimensional chaos can occur even in a game as simple as the one we study here. In more complicated games with higher dimensional chaos, we expect that it becomes more common."[48]

---

[37] Ibid. p. 3
[38] For a dynamic visual presentation of these graphics, see http://www.santafe.edu/~jdf/papers/rps.pdf
[39] Ibid., $\epsilon_x + \epsilon y < 0$ (e.g. $\epsilon_x = -0.1$, $\epsilon_y = 0.5$) p. 3
[40] Ibid., In contrast, the single population replicator model, the SAF team found that the time spent on pure strategies varied exponentially with time. (p. 3)
[41] Ibid., p. 4, Figure 2.
[42] Ibid. No. 41, see also p. 4, Figure 3. The authors describe this behavior as similar to that in Figure 2, but the time spent near each pure orbit varies irregularly. "The orbit is an infinitely persistent chaotic behavior, a dynamic interest in its own right, (that to our knowledge does not occur in the single population replicator model). p. 4
[43] Ibid., No. 3, p. 4
[44] (a) Multiple iteration games too are not without their counter-intuitive elements. Perhaps the most notable example of this is Robert Axelrod's multi-generation computer tournament, where the simple "tit-for-tat" strategy outperformed a bevy of professional game theorists invited to present strategies for the tournament (Ibid. No. 20). More recently, several authors have explored the dynamics of both simple and complex multiple iteration games under various assumptions of rationality and with both local and global optimization. Such treatments of multiple iteration games have been undertaken by (b) Cohen, Riolo and Axelrod, "Emergence of Social Organization in the Prisoner's Dilemma", Santa Fe Institute, Working Papers, 99-01-002 as well as (c) David Kane in "Local Hill climbing on an Economic Landscape", Santa Fe Institute, Working Papers, 96-08-065. (d) Vince Darley, gives a particularly interesting mathematical treatment of what he describes as autonomous "dispersed agents" in his chapter four, "Nonlinear Models of Intermittent Dynamics" of his Harvard University Masters Thesis, Towards a Theory of Autonomous, Optimising Agents, (1999). (e) Finally, Ricard V. Sole, Suzanna C. Manrubia, Michael Benton, Stuart Kauffman, and Per Bak, in "Criticality and Scaling in Evolutionary Ecology" examine the fractal nature of evolutionary processes and their occurrence in ecological systems. The origin of these dynamical features is investigated and self-organized criticality suggested as a common feature, leading to power-law predictions of extinction events and spatio-temporal diversity, Santa Fe Institute, Working Papers, 97-12-086.

## The Complexity of Experience—What does it mean? What is it Good For?

So where does that leave the reader after fighting through the complexity of Hamiltonian dynamics, heteroclinic orbits, Lyapunov spectra, conserved phase space across pairwise conjugate coordinates, and four dimensional simplexes? Should we simply throw up our hands in frustration having discovered that many of the simplest game theoretic characterizations of strategy and behavior involve calculationally intractable problems? A look at some of the other work of the SAF team[49] as well as recent work in quantum computing suggests otherwise.[50] There are a variety of methods for handling complex, multidimensional conceptual and calculational problems which have emerged from recent studies.[51] Even calculational intractability has particular defining characteristics which may prove to be less intractable under other conditions. Also the work of Farmer and others on modeling through the study of "local rules of behavior" may lead to ways in which even computationally intractable simulations can be employed at a local level of study. For example, we have the SAF team's argument about chaos and self-consistency performing as a kind of strategic "protective coloration" in a competitive bidding environment.

Certainly, it is a discovery of major scientific importance that complex chaotic dynamics can be found in even the simplest of systems. Taken from the standpoint of competitive strategy, this finding should provide encouragement for researchers to discover applications and to build strategies which take advantage of these newly discovered systems properties.

In the larger arena of competitive strategy in general, the SAF team's findings have profound implications for the discovery of previously unrecognized or even unsuspected characteristics in agent-based models.[52] Their findings suggest that, depending on the particulars of the given situation, there may be a variety of strategies for exploiting this new set of dynamics, particularly when such dynamics are not easily visible (i.e. low market transparency, high information asymmetries and/or high transaction costs.) or are poorly understood by competitors. Similarly, the explicit recognition of the structure of these new dynamics may also allow the implementation of formal systems to enable more sophisticated players to avoid wasting time or allocating resources to the process of attempting to construct regression driven trend models or linear predictive strategies in areas which are too complex to predict with such tools, inherently impossible to predict or theoretically possible to predict but which are currently calculationally intractable.

---

[45] In fairness to the authors, this is their ex-post ante analysis, and while it is true that retrospectively chaos introduces complex problems into the learning process of the opponent if the process were actually self-conscious, then instead of arbitrary or chaotic learning trajectories, all relatively skilled agents would proceed rapidly to the mixed strategy Nash Equilibrium simply on an intuitive basis, although the empirical evidence is quite clear that this is NOT what happens even in the simplest games.
[46] Ibid. p. 4
[47] Ibid. p. 4
[48] Ibid., The authors also note that "Many economists have noted the lack of any compelling account of how agents might learn to play a Nash Equilibrium. Our results strongly reinforce this concern, in a game simple enough for children to play. The fact that chaos can occur in learning such a simple game indicates that one should use caution in assuming that real people will learn to play a game according to a Nash Equilibrium strategy." (p. 4)
[49] Sato, Yuzuru, Makoto Taiji and Takashi Ikegami, "Computation with Switching Map Systems: Nonlinearity and Computational Complexity", Santa Fe Institute Working Papers, 01-12-083, also appearing in Journal of Universal Computer Science, Volume 6, No. 9, 2000.
[50] (a) Christopher Moore, and Martin Nilsson, "Parallel Quantum Computation and Quantum Codes", Santa Fe Institute Working Papers, 98-08-070, appearing in SIAM Journal on Computing, Volume 31, No. 3, 2001, pp. 799-815. (b) Path Integration on a Quantum Computer" by Joseph F. Traub and Henryk Wozniakowksi, Santa Fe Institute, Working Paper, 01-10-55..
[51] See (a) Terraneo, M. and Shepelyansky, D.L. (2004) "Dynamical Localization and Repeated Measurements in a Quantum Computation Process" Phys. Rev. Lett. 92, 037902 (2004); also (b) Benenti, G., Casati, G., Montangero, S., and Shepelyansky, D.L., (2003) Statistical properties of eigenvalues for an operating quantum computer with static imperfections" European Physical Journal D 22, 2, 285-293 (2003).

---

[52] See (a) Farmer, J. Doyne, Toward Agent-Based Models for Investment." in Developments in Quantitative Investment Models, AIMR, 2001. See also (b) Farmer, J. D., Patelli, P., and Zovko, I. "The predictive power of zero intelligence Models in Financial Markets." Los Alamos National Laboratory Condensed Matter archive 0309233 (2003).

# Appendix I: The Minority Game (Agent Based Modeling)[53]

The minority game represents the opposite end of the spectrum. Despite its simplicity, it displays some rich behavior. While the connection to markets is only metaphorical, its behavior hints at the problems with the traditional views of efficiency and equilibrium. The minority game was originally motivated by Brian Arthur's El Farol problem.[54] El Farol is a bar in Santa Fe, near the original site of the Santa Fe Institute, which in the old days was a popular hangout for SFI denizens. In the El Farol problem, a fixed number of agents face the question of whether or not to attend the bar. If the bar is not crowded, as measured by a threshold on the total number of agents, an agent wins if he or she decides to attend. If the bar is too crowded, the agent wins by staying home. Agents make decisions based on the recent record of total attendance at the bar. This problem is like a market in that each agent tries to forecast the behavior of the aggregate and that no outcome makes everyone happy.

The minority game introduced by Damien Challet and Yi-Cheng Zhang is a more specific formulation of the El Farol problem. At each timestep, N agents choose between two possibilities (for example, A and B). A historical record is kept of the number of agents choosing A; because N is fixed, this automatically determines the number who chose B. The only information made public is the most popular choice. A given time step is labeled "0" if choice A is more popular and "1" if choice B is more popular. The agents' strategies are lookup tables whose inputs are based on the binary historical record for the previous m timesteps. Strategies can be constructed at random by simply assigning random outputs to each input (see Table A).

Each agent has s possible strategies, and at any given time plays the strategy that has been most successful up until that point in time. The ability to test multiple strategies and use the best strategy provides a simple learning mechanism. This learning is somewhat effective—for example, asymptotically A is chosen 50% of the time. But because there is no choice that satisfies everyone—indeed, no choice that satisfies the majority of the participants—there is a limit to what learning can achieve for the group as a whole.

Table A.

| Input | Output |
|-------|--------|
| 0 0   | 1      |
| 0 1   | 0      |
| 1 0   | 0      |
| 1 1   | 1      |

Example of a strategy for the minority game. The input is based on the attendance record for the m previous time-steps, 0 or 1, corresponding to which choice was most popular. In this case m = 2. The output of the strategy is its choice (0 or 1). Outputs are assigned t random.

When $s > 1$, the sequence of 0s and 1s corresponding to the attendance record is aperiodic. This is driven by switching between strategies. The set of active strategies continues to change even though the total pool of strategies is fixed. For a given number of agents, for small m the game is efficient, in that prediction is impossible, but when m is large, this is no longer the case. In the limit $N \to \infty$, as m increases there is a sharp transition between the efficient and the inefficient regime.

The standard deviation of the historical attendance record, $\sigma$, provides an interesting measure of the average utility. Assume that each agent satisfies his or her utility function by making the minority choice. The average utility is highest when the two choices are almost equally popular. For example, with 101 agents the maximum utility is achieved if 50 agents make one choice and 51 the other. However, it is impossible to achieve this state consistently. There are fluctuations around the optimal attendance level, lowering the average utility. As m increases, $\sigma$ exhibits interesting behavior, starting out at a maximum, decreasing to a minimum, and then rising to obtain an asymptotic value in the limit as $m \to \infty$. The minimum occurs at the transition between the efficient and inefficient regimes. The distinction between the efficient and inefficient regimes arises from the change in the size of the pool of strategies present in the population, relative to the total number of possible strategies. The size of the pool of strategies is *sN*. The number of possible strategies is $2^{2^m}$, which grows extremely rapidly with m. For example, for m = 2 there are 16 possible strategies, for m = 5 there are roughly 4 billion, and for m =10 there are more than $10^{300}$—far exceeding the number of elementary particles in the universe. In contrast, with s = 2 and N = 100, there are only 200 strategies actually present in the pool. For low m, when the space of strategies is well-covered, the conditional probability for a given transition is the same for all histories—there are no patterns of length m. But when m is larger, so that the strategies are only sparsely filling the space of possibilities, patterns remain. We can interpret this as meaning that the market is efficient for small m and inefficient for large m. The El Farol problem and minority game is a simple game with no solution that can satisfy everyone. This is analogous to a market where not everyone profits on any given trade. *Studies of the minority game suggest that the long-term behavior is aperiodic: the aggregate behavior continues to fluctuate. In contrast to the standard view in economics, such fluctuations occur even in the absence of any new external information.*

---

[53] Excerpted from "J. Doyne Farmer, "Physicists Attempt to Scale the Ivory Towers of Finance", Computing in Science and Engineering, November-December, 1999.
[54] Ibid.

# Appendix II: The Dynamics of a Generalized Rock-Paper-Scissors Game[55]

The generalized form of the game is:

$$A = \begin{bmatrix} \epsilon x & -1 & 1 \\ 1 & \epsilon x & -1 \\ -1 & 1 & \epsilon x \end{bmatrix} \quad B = \begin{bmatrix} \epsilon y & -1 & 1 \\ 1 & \epsilon y & -1 \\ -1 & 1 & \epsilon y \end{bmatrix}$$

Where $-1 \leq \epsilon x \leq 1$ and $-1 \leq \epsilon y \leq 1$ are the payoffs when there is a tie.[56] We have placed the columns in the order "rock", "paper", and "scissors". For example, reading down the first column of **A**, in the case that the opponent plays "rock", we see that the payoff for using the pure strategy "rock" is $\epsilon x$, "paper" is 1, and "scissors" is $-1$.

The rock, paper scissors game exemplifies a fundamental and important class of games where no strategy is dominant and no pure Nash Equilibrium exists (any pure strategy is vulnerable to another).[57]

---

[55] From Sato, Yuzuru, Akiyama, Eizo and J. Doyne Farmer, "Chaos in a Simple Two Person Learning Game", Santa Fe Institute Working Papers, 01-09-049.
[56] Ibid., The authors place these bounds on $\epsilon$ because when they are violated, the behavior under ties dominates, and this more closely resembles a "matching-pennies" type game with three strategies (and hence a different game model).
[57] Ibid.